\documentstyle[sprocl,epsf]{article}

\begin{document}

\title{NEUTRINOS FROM THE NEXT GALACTIC SUPERNOVA}
\author{J. F. BEACOM}
\address{California Institute of Technology, Physics 161-33\\
1200 E. California Blvd., Pasadena, CA 91125 USA\\
E-mail: beacom@citnp.caltech.edu}

\maketitle\abstracts{The next core-collapse supernova in our Galaxy
will be a spectacular event, with some $10^4$ neutrino detections in
total expected among several detectors.  This data will allow
unprecedented tests of neutrino properties and new opportunities in
astrophysics.  In this paper, I focus on two main topics: (1)
Measurement of the $\nu_\mu$ and $\nu_\tau$ masses by time-of-flight,
with an emphasis on introducing as little supernova model dependence
as possible, and (2) Methods for locating a supernova by its neutrinos
in advance of the light, which may allow improved astronomical
observations.  In the latter, I also discuss the recent result that
the positrons from $\bar{\nu}_e + p \rightarrow e^+ + n$ are {\it not}
isotropically emitted, as commonly thought.}


\section{Supernova Neutrino Mass Measurements}


\subsection{Current Knowledge About Neutrino Masses}

What is known about the neutrino masses, in terms of direct
experimental evidence?  After several decades of experiments on the
tritium beta spectrum, the published limit on the mass of the
$\bar{\nu}_e$ (and hence also $\nu_e$) is now about 3
eV~\cite{tritium}.  These results are still attended by some
controversy, due to apparent systematic errors near the endpoint
(discussed in detail at a recent workshop~\cite{INT} at the Institute
for Nuclear Theory in Seattle).  Much less is known about the
$\nu_\mu$ mass ($< 170$ keV from $\pi$ decay)~\cite{Assamagan}, and
the $\nu_\tau$ mass ($< 18$ MeV from $\tau$ decay)~\cite{Barate}.  It
would be a great advance if the latter two limits could be improved to
the eV or tens of eV level, and it seems unlikely that any terrestrial
experiment could do that.

There are two indirect arguments which {\it may} restrict the
$\nu_\mu$ and $\nu_\tau$ masses.  The first says that the predicted
background of relic neutrinos ($\approx 100/{\rm cm}^3$) should not
overclose the universe.  This leads to a limit on the sum of the
neutrino masses of about 100 eV in the most conservative case, and
about 10 eV if reasonable values for the non-baryonic dark matter
density and the Hubble constant are used~\cite{Fisher}.  The second is
based on the recent evidence for neutrino oscillations (sensitive only
to {\it differences} of neutrino masses).  For three neutrino flavors,
there are two independent mass differences, which can be chosen to
explain the solar and atmospheric neutrino data.  The overall scale
can be fixed by using the limit from tritium beta decay, and since the
observed mass differences are small, then all masses are below a few
eV~\cite{Barger}.

These indirect arguments sound compelling, but they are not hard to
evade.  The cosmological bound doesn't apply if the neutrinos are
allowed to decay.  There are many such models for heavy $\nu_\tau$
masses, motivated by particle physics~\cite{heavy1} or
astrophysics~\cite{heavy2}.  The oscillation bound doesn't apply if
there are more free mass differences than there are positive signals
of neutrino oscillations.  For example, suppose the atmospheric
neutrinos are undergoing maximal $\nu_\mu \leftrightarrow \nu_\tau$
mixing, LSND is ruled out, and the solar neutrinos are undergoing
$\nu_e \leftrightarrow \nu_{sterile}$ mixing.  Then nothing prevents
the $\nu_2$ and $\nu_3$ masses from having a tiny splitting near 50 or
even 500 eV, and leaving the $\nu_1$ mass below a few eV.

It may be possible in the future to make new astrophysical tests of
the neutrino masses, using data on large-scale
structure~\cite{Hu1,Fukugita}, the cosmic microwave
background~\cite{Lopez}, weak lensing~\cite{Cooray}, and the
Lyman-alpha forest~\cite{Hu2}.  While these techniques claim
sensitivities of order 1 - 10 eV for the sum of the neutrino masses,
none are based on direct detection of neutrinos, and hence may be
vulnerable to uncertainties in their assumptions.

Thus while indirect evidence will be valuable, there is still a need
for a {\it direct} measurement of the $\nu_\mu$ and $\nu_\tau$ masses
(or, in the presence of mixing, the masses of the heavy mass
eigenstates).


\subsection{Supernova Neutrino Emission}

When the core of a large star ($M \ge 8 M_{\odot}$) runs out of
nuclear fuel, it collapses, with a change in the gravitational binding
energy of about $3 \times 10^{53}$ ergs.  This huge amount of energy
must be removed from the proto-neutron star, but the high density
prevents the escape of any radiation.  Inside of the hot proto-neutron
star, neutrino-antineutrino pairs of all flavors are produced.
Despite their weak interactions, even the neutrinos are trapped and
diffuse out over several seconds, in the end carrying away about 99\%
of the supernova energy.  (I have neglected the flux of $\nu_e$
neutrinos necessary to change the core from $N \simeq Z$ nuclei into a
neutron star, because the total energy release of this flux is of
order 1\% of the pair emission phase).

When the neutrinos are about one mean free path from the edge, they
escape freely, with a thermal spectrum (approximately Fermi-Dirac)
characteristic of the surface of last scattering.  Because different
flavors have different interactions with the matter, and because the
neutron star temperature is decreasing with increasing radius, the
neutrino decoupling temperatures are different.  The $\nu_\mu$ and
$\nu_\tau$ neutrinos and their antiparticles have a temperature of
about 8 MeV, the $\bar{\nu}_e$ neutrinos about 5 MeV, and the $\nu_e$
neutrinos about 3.5 MeV.  Equivalently, the average energies are about
$\langle E \rangle \simeq$ 25 MeV, $\simeq 16$ MeV, and $\simeq 11$
MeV, respectively.  The luminosities of the different neutrino flavors
are approximately equal at all times.  The neutrino luminosity rises
quickly over 0.1 s or less, and then falls off over several seconds.
The SN1987A data can be reasonably fit to a decaying exponential with
time constant $\tau$ = 3 s.  The detailed form of the neutrino
luminosity used below is less important than the general shape
features and their characteristic durations.

The estimated core-collapse supernova rate in the Galaxy is about 3
times per century~\cite{SNrate}.  The present neutrino detectors can
easily observe a supernova anywhere in the Galaxy or its immediate
companions (e.g., the Magellanic Clouds).  Unfortunately, the present
detectors do not have large enough volumes to observe a supernova in
even the nearest galaxy (Andromeda, about 700 kpc away).


\subsection{Time-of-Flight Concept}

Even a tiny neutrino mass will make the velocity slightly less than
for a massless neutrino, and over the large distance to a supernova
will cause a measurable delay in the arrival time.  A neutrino with a
mass $m$ (in eV) and energy $E$ (in MeV) will experience an
energy-dependent delay (in s) relative to a massless neutrino in
traveling over a distance D (in 10 kpc, approximately the distance to
the Galactic center) of
\begin{equation}
\Delta t(E) = 0.515
\left(\frac{m}{E}\right)^2 D\,,
\label{eq:delay}
\end{equation}
where only the lowest order in the small mass has been kept.  A
distance of 10 kpc corresponds to a travel time of about $10^{12}$ s,
and the delays of interest are less than 1 s.

If the neutrino mass is nonzero, lower-energy neutrinos will arrive
later, leading to a correlation between neutrino energy and arrival
time.  This is exploitable in some of the charged-current detection
reactions for $\nu_e$ and $\bar{\nu}_e$ since the neutrino energy can
be reasonably measured.  Using this idea, the next supernova will
allow sensitivity to a $\bar{\nu}_e$ mass down to about 3
eV~\cite{Totani}.  However, the terrestrial experiments now
claim~\cite{tritium} to rule out a mass that large, so I ignore the
$\nu_e$ mass in describing the supernova neutrino signal.

For measuring the masses of the $\nu_\mu$ and $\nu_\tau$ neutrinos,
one must realize that only neutral-current detection reactions are
possible (the charged-current thresholds are too high).  Since the
neutrino energy is {\it not} measured in neutral-current interactions,
an energy-time correlation technique cannot be used (the incoming
neutrino energy is not determined since a complete kinematic
reconstruction of the reaction products is not possible).

Instead, the strategy for measuring the $\nu_\tau$ mass is to look at
the difference in time-of-flight between the neutral-current events
(mostly $\nu_\mu$,$\nu_\tau$,$\bar{\nu}_\mu$, and $\bar{\nu}_\tau$,
because of the higher temperature of those flavors) and the
charged-current events (just $\nu_e$ and $\bar{\nu}_e$).  We assume
that the $\nu_\mu$ is massless and will ask what limit can be placed
on the $\nu_\tau$ mass (other cases will be discussed below).  There
are three major complications to a simple application of
Eq.~(\ref{eq:delay}): (i) The neutrino energies are not fixed, but are
characterized by spectra; (ii) The neutrino pulse has a long intrinsic
duration of about 10 s, as observed for SN1987A; and (iii) The
statistics are finite.  In some early work~\cite{oldtime}, the
duration of the pulse was thought to be of order 100 milliseconds,
nearly a delta-function.  Here (considering the case of the smallest
detectable mass), the delay is much less than the width of the pulse,
which is what makes the problem much more difficult (the large-mass
case is covered elsewhere~\cite{SKpaper,SNOpaper}).  Given that the
energy-time correlation method cannot be used, we must turn to
considerations of the shape of the event rate as a function of time.


\subsection{Neutrino Scattering Rate}

For a time-independent spectrum $f(E)$, and luminosity $L(t_i)$,
the double-differential number distribution at the source is
\begin{equation}
\frac{d^2 N_\nu}{dE_\nu dt_i} = 
f(E_\nu) \frac{L(t_i)}{\langle E_\nu \rangle}\,.
\end{equation}
The neutrino spectrum and average energy $\langle E_\nu \rangle$ are
different for each flavor, but the luminosities $L(t_i)$ of the
different flavors are roughly equal~\cite{Janka}, so that each flavor
carries away about 1/6 of the total binding energy $E_B$.  For a
massive neutrino, the double-differential number distribution at the
detector is
\begin{equation}
\frac{d^2 N_\nu}{dE_\nu dt} = 
f(E_\nu) \frac{L(t - \Delta t(E_\nu))}{\langle E_\nu \rangle}
\end{equation}
and the scattering rate for a given reaction is
\begin{equation}
\frac{dN_{sc}}{dt} = 
C \int dE_\nu f(E_\nu) 
\left[\frac{\sigma(E_\nu)}{10^{-42} {\rm cm}^2}\right]
\left[\frac{L(t - \Delta t(E_\nu))}{E_B/6}\right]\,.
\end{equation}
The overall constant for an H$_2$O target is
\begin{equation}
C = 9.21
\left[\frac{E_B}{10^{53} {\rm\ ergs}}\right]
\left[\frac{1 {\rm\ MeV}}{T}\right]
\left[\frac{10 {\rm\ kpc}}{D}\right]^2
\left[\frac{{\rm det.\ mass}}{1 {\rm\ kton}}\right]
\,n\,,
\label{eq:C}
\end{equation}
where $n$ is the number of targets per molecule for the given
reaction.  For a D$_2$O target, the leading factor becomes 8.28
instead of 9.21.

For a massless neutrino, $\Delta t(E_\nu) = 0$, so the luminosity
comes outside of the integral as $L(t)$ and completely specifies the
time dependence.  For a massive neutrino, the time dependence from the
luminosity is modified in an energy-dependent way by the mass effects,
as above.

\begin{table}[t]
\caption{Calculated numbers of events expected in SK (32 kton H$_2$O)
with a 5 MeV threshold and a supernova at 10 kpc.  The other
parameters (e.g., neutrino spectrum temperatures) are given in the
text.  In rows with two reactions listed, the number of events is the
total for both.  The second row is a subset of the first row that is
an irreducible background to the reactions in the third and fourth
rows~\protect\cite{oxy}. The detected final-state particles are 
$e^+,e^-,\gamma$.}
\vspace{0.2cm}
\begin{center}
\begin{tabular}{|l|l|}
\hline
reaction & number of events \\
\hline\hline
$\bar{\nu}_e + p \rightarrow e^+ + n$ & 8300 \\
\hline
$\bar{\nu}_e + p \rightarrow e^+ + n$
 \phantom{aa} ($E_{e^+} \leq 10$ MeV) & 530 \\
\hline
$\nu_\mu + ^{16}{\rm O} \rightarrow \nu_\mu + \gamma + X$ & 355 \\
$\bar{\nu}_\mu + ^{16}{\rm O} \rightarrow \bar{\nu}_\mu + \gamma + X$ & \\
\hline
$\nu_\tau + ^{16}{\rm O} \rightarrow \nu_\tau + \gamma + X$ & 355 \\
$\bar{\nu}_\tau + ^{16}{\rm O} \rightarrow \bar{\nu}_\tau + \gamma + X$ & \\
\hline
$\nu_e + e^- \rightarrow \nu_e + e^-$ & 200 \\
$\bar{\nu}_e + e^- \rightarrow \bar{\nu}_e + e^-$ & \\
\hline
$\nu_\mu + e^- \rightarrow \nu_\mu + e^-$ & 60 \\
$\bar{\nu}_\mu + e^- \rightarrow \bar{\nu}_\mu + e^-$ & \\
\hline
$\nu_\tau + e^- \rightarrow \nu_\tau + e^-$ & 60 \\
$\bar{\nu}_\tau + e^- \rightarrow \bar{\nu}_\tau + e^-$ & \\
\hline
\end{tabular}
\end{center}
\end{table}


\subsection{Separation of the Massive Signal}

The scattering rates of the massless neutrinos ($\nu_e$ or
$\bar{\nu}_e$) can then be used to measure the shape of the luminosity
as a function of time.  The scattering rate for the massive neutrinos
($\nu_\tau$) can be tested for additional time dependence due to a
mass.  We define two rates: a Reference $R(t)$ containing only
massless events, and a Signal $S(t)$ containing some fraction of
massive events.

The Reference $R(t)$ can be formed in various ways, for example from
the charged-current reaction $\bar{\nu}_e + p \rightarrow e^+ + n$ in
the light water of SK or SNO.  (The numbers of events expected for all
reactions are given in Table~I for SK and Table~II for SNO).

The Signal $S(t)$ can be based on various neutral-current
reactions~\cite{SKpaper,SNOpaper,Cadonati}, though here I will focus
on the results for SNO~\cite{SNOpaper}.  Thus the primary component of
the Signal $S(t)$ are the 485 neutral-current events on deuterons.
With the hierarchy of temperatures assumed here, these events are 18\%
($\nu_e + \bar{\nu}_e$), 41\% ($\nu_\mu + \bar{\nu}_\mu$), and 41\%
($\nu_\tau + \bar{\nu}_\tau$).  The flavors of the neutral-current
events of course cannot be distinguished.  Under our assumption that
only $\nu_\tau$ is massive, there is already some unavoidable dilution
of $S(t)$.

\begin{table}[t]
\caption{Calculated numbers of events expected in SNO for a supernova
at 10 kpc.  The other parameters (e.g., neutrino spectrum temperatures)
are given in the text.  In rows with two reactions listed, the number of
events is the total for both.  The middle column indicates the detected
particle(s).  The notation $\nu$ indicates the sum of $\nu_e$, $\nu_\mu$,
and $\nu_\tau$, though they do not contribute equally to a given reaction, 
and $X$ indicates either $n + ^{15}$O or $p + ^{15}$N.}
\vspace{0.2cm}
\begin{center}
\begin{tabular}{|l|r|r|}
\hline
\multicolumn{3}{|c|}{Events in 1 kton D$_2$O}\\
\hline\hline
$\nu + d \rightarrow \nu + p + n$ & $n$ & 485 \\
$\bar{\nu} + d \rightarrow \bar{\nu} + p + n$ & $n$ & \\
\hline
$\nu_e + d \rightarrow e^- + p + p$ & $e^-$ & 160 \\
$\bar{\nu}_e + d \rightarrow e^+ + n + n$ & $e^+nn$ & \\
\hline
$\nu + ^{16}{\rm O} \rightarrow \nu + \gamma + X$ & 
$\gamma$, $\gamma n$ & 20 \\
$\bar{\nu} + ^{16}{\rm O} \rightarrow \bar{\nu} + \gamma + X$ &
$\gamma$, $\gamma n$ & \\
\hline
$\nu + ^{16}{\rm O} \rightarrow \nu + n + ^{15}{\rm O}$ & $n$ & 15 \\
$\bar{\nu} + ^{16}{\rm O} \rightarrow \bar{\nu} + n + ^{15}{\rm O}$ & $n$ & \\
\hline
$\nu + e^- \rightarrow \nu + e^-$ & $e^-$ & 10 \\
$\bar{\nu} + e^- \rightarrow \bar{\nu} + e^-$ & $e^-$ & \\
\hline
\multicolumn{3}{|c|}{Events in 1.4 kton H$_2$O}\\
\hline\hline
$\bar{\nu}_e + p \rightarrow e^+ + n$ & $e^+$ & 365 \\
\hline
$\nu + ^{16}{\rm O} \rightarrow \nu + \gamma + X$ & $\gamma$ & 30 \\
$\bar{\nu} + ^{16}{\rm O} \rightarrow \bar{\nu} + \gamma + X$ & $\gamma$ & \\
\hline
$\nu + e^- \rightarrow \nu + e^-$ & $e^-$ & 15 \\
$\bar{\nu} + e^- \rightarrow \bar{\nu} + e^-$ & $e^-$ & \\
\hline
\end{tabular}
\end{center}
\end{table}

In Fig.~1, $S(t)$ is shown under different assumptions about the
$\nu_\tau$ mass.  The shape of $R(t)$ is exactly that of $S(t)$ when
$m_{\nu_\tau} = 0$, though the number of events in $R(t)$ will be
different.  The rates $R(t)$ and $S(t)$ will be measured with finite
statistics, so it is possible for statistical fluctuations to obscure
the effects of a mass when there is one, or to fake the effects when
there is not.  We determine the mass sensitivity in the presence of
the statistical fluctuations by Monte Carlo modeling.  We use the
Monte Carlo to generate representative statistical instances of the
theoretical forms of $R(t)$ and $S(t)$, so that each run represents
one supernova as seen in SNO.  The best model-independent test of a
$\nu_\tau$ mass seems to be a test of the average arrival time
$\langle t \rangle$.  Any massive component in $S(t)$ will always
increase $\langle t \rangle$, up to statistical fluctuations.

\begin{figure}
\epsfxsize=4.5in \epsfbox{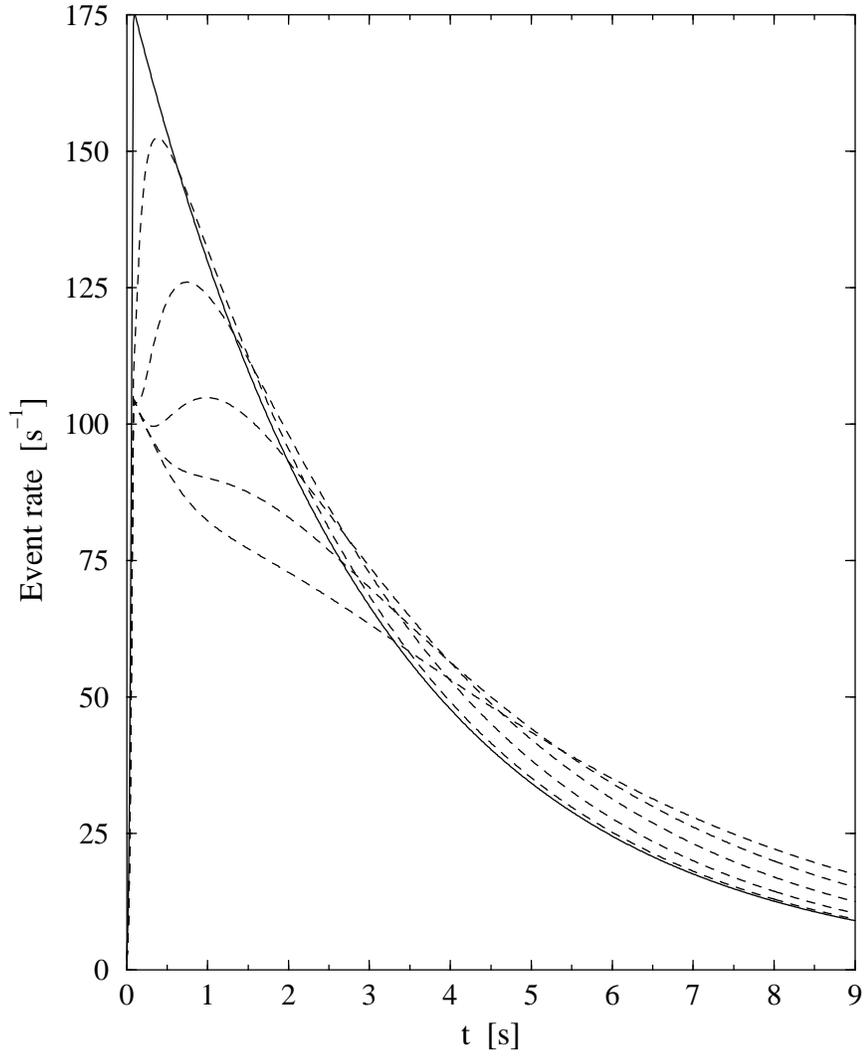}
\caption{The expected event rate for the Signal $S(t)$ at SNO in the
absence of fluctuations for different $\nu_\tau$ masses, as follows:
solid line, 0 eV; dashed lines, in order of decreasing height: 20, 40,
60, 80, 100 eV.  Of 535 total events, 100 are massless ($\nu_e +
\bar{\nu}_e$), 217.5 are massless ($\nu_\mu + \bar{\nu}_\mu$), and
217.5 are massive ($\nu_\tau + \bar{\nu}_\tau$).  These totals count
events at all times; in the figure, only those with $t \le 9$ s are
shown.}
\end{figure}


\subsection{$\langle t \rangle$ Analysis}

Given the Reference $R(t)$ (i.e., the charged-current events), the
average arrival time is defined as
\begin{equation}
\langle t \rangle_R = \frac{\sum_k t_k}{\sum_k 1}\,,
\end{equation}
where the sum is over events in the Reference.  The effect of the
finite number of counts $N_R$ in $R(t)$ is to give $\langle t \rangle_R$
a statistical error:
\begin{equation}
\delta\left(\langle t \rangle_R\right) =
\frac{\sqrt{\langle t^2 \rangle_R - \langle t \rangle^2_R}}{\sqrt{N_R}}\,.
\label{eq:tRerror}
\end{equation}
For a purely exponential luminosity, $\langle t \rangle_R =
\sqrt{\langle t^2 \rangle_R - \langle t \rangle^2_R} = \tau \simeq 3$ s.
Given the Signal $S(t)$ (i.e., the neutral-current events), the
average arrival time $\langle t \rangle_S$ and its error
$\delta\left(\langle t \rangle_S\right)$ are defined similarly.
The signal of a mass is that the measured value of $\langle
t \rangle_S - \langle t \rangle_R$ is greater than zero with
statistical significance.

Using the Monte Carlo, we analyzed $10^4$ simulated supernova data
sets for a range of $\nu_\tau$ masses.  For each data set, $\langle t
\rangle_S - \langle t \rangle_R$ was calculated and its value
histogrammed.  These histograms are shown in the upper panel of Fig.~2
for a few representative masses.  (Note that the number of Monte Carlo
runs only affects how smoothly these histograms are filled out, and
not their width or placement.)  These distributions are characterized
by their central point and their width, using the 10\%, 50\%, and 90\%
confidence levels.  That is, for each mass we determined the values of
$\langle t \rangle_S - \langle t \rangle_R$ such that a given
percentage of the Monte Carlo runs yielded a value of $\langle t
\rangle_S - \langle t \rangle_R$ less than that value.  With these
three numbers, we can characterize the results of complete runs with
many masses much more compactly, as shown in the lower panel of
Fig.~2.  Given an experimentally determined value of $\langle t
\rangle_S - \langle t \rangle_R$, one can read off the range of masses
that would have been likely (at these confidence levels) to have given
such a value of $\langle t \rangle_S - \langle t \rangle_R$ in one
experiment.  From the lower panel of Fig.~2, we see that SNO is
sensitive to a $\nu_\tau$ mass down to about 30 eV if the SK $R(t)$ is
used, and down to about 35 eV if the SNO $R(t)$ is used.

We also investigated the dispersion of the event rate in time as a
measure of the mass.  A mass alone causes a delay, but a mass and an
energy spectrum also causes dispersion (as does the separation of the
massive and massless portions of the Signal).  We defined the
dispersion as the change in the width $\sqrt{\langle t^2 \rangle_S -
\langle t \rangle^2_S} - \sqrt{\langle t^2 \rangle_R - \langle t
\rangle^2_R}$.  We found that the dispersion was not statistically
significant until the mass was of order 80 eV or so, i.e., a plot very
similar to Fig.~2 can be formed for the width, and the error band is
much wider.  Since the dispersion is irrelevant, the average delay is
well-characterized by a single energy, which for SNO is $E_c \simeq
32$ MeV.

\begin{figure}
\epsfxsize=4.5in \epsfbox{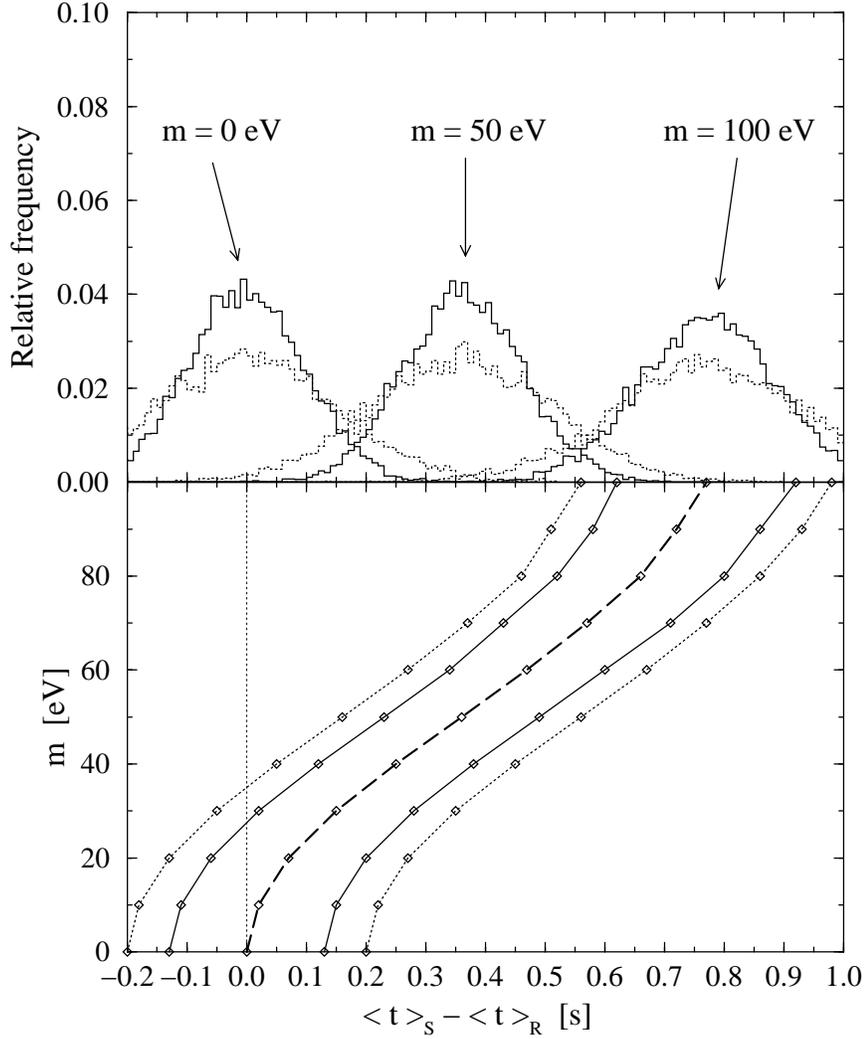}
\caption{The results of the $\langle t \rangle$ analysis for a massive
$\nu_\tau$, using the Signal $S(t)$ from SNO defined in the text.  In
the upper panel, the relative frequencies of various $\langle t
\rangle_S - \langle t \rangle_R$ values are shown for a few example
masses.  The solid line is for the results using the SK Reference
$R(t)$, and the dotted line for the results using the SNO $R(t)$.  In
the lower panel, the range of masses corresponding to a given $\langle
t \rangle_S - \langle t \rangle_R$ is shown.  The dashed line is the
50\% confidence level.  The upper and lower solid lines are the 10\%
and 90\% confidence levels, respectively, for the results with the SK
$R(t)$.  The dotted lines are the same for the results with the SNO
$R(t)$.}
\end{figure}


\subsection{Analytic Results}

Another nice feature of the proposed $\langle t \rangle$ analysis,
besides its simplicity, is that good estimates can be made
analytically.  The characteristic delay is
\begin{equation}
\langle t \rangle_S - \langle t \rangle_R \simeq
{\rm frac}(m > 0) \times 0.515 \left(\frac{m}{E_c}\right)^2 D\,,
\end{equation}
where ${\rm frac}(m > 0)$ is the fraction (about 1/2) of massive
events in the neutral-current Signal.  The above numerical analysis
shows that the mass effects occur most significantly in the delay and
not the width (are characterized with a single energy, and negligible
dispersion).  That characteristic energy $E_c$ is roughly at the peak
of $f(E)\sigma(E)$ (see Fig.~3).  This formula for the first moment is
always true; the point of the Monte Carlo analysis was to show that
the other moments are not changed, so this completely describes the
data.

If the cross section $\sigma(E_\nu)$ depends on energy as
$E_\nu^{\alpha}$ ($\alpha \sim 2$ for $\nu + d$), then the
characteristic energy $E_c \sim (2+\alpha)T$ and the
thermally-averaged cross section is proportional to $T^{\alpha}$,
where $T$ is the $\nu_\mu$ and $\nu_\tau$ temperature.  For $\alpha =
2$, and a neutron detection efficiency $\epsilon_n$, the following
scaling relations hold:
\begin{equation}
\langle t \rangle_S - \langle t \rangle_R \sim
\left(\frac{m}{T}\right)^2 D\,.
\end{equation}
\begin{equation}
N_S \sim \frac{1}{T} \frac{1}{D^2}\,T^2 \epsilon_n
\sim \frac{T \epsilon_n}{D^2}\,,
\end{equation}
\begin{equation}
\delta\left(\langle t \rangle_S - \langle t \rangle_R\right) \sim
\frac{\tau}{\sqrt{N_S}} \sim \frac{\tau D}{\sqrt{T}\sqrt{\epsilon_n}}\,.
\end{equation}
For a non-exponential luminosity, $\tau$ is more generally the width
of the pulse.  If zero delay is measured, then the mass limit is
determined by the largest positive delay that could have fluctuated to
zero, and thus
\begin{equation}
m_{limit} \sim T^{3/4}\, \tau^{1/2} \epsilon_n^{-1/4}\,.
\label{eq:mlimit}
\end{equation}
Note that is {\it independent} of $D$ for any supernova distance in
the Galaxy (i.e., as long as there is a reasonable number of counts).
Note that the mass limit scales with the {\it fourth} root of the
number of events (via the efficiency $\epsilon_n$).

Besides the scaling, the value of $m_{lim}$ at fixed values of these
quantities can also be found easily~\cite{SNOpaper}.  The analytic
results above are in excellent agreement with the full numerical
results~\cite{SNOpaper}.  Over {\it reasonable} ranges of the input
parameters, the results hardly change.  However, some estimates of the
$\nu_\tau$ mass sensitivity for proposed detectors~\cite{LAND,OMNIS}
assumed a very short duration of the supernova ($\tau \simeq 0.5$ s).
If such a short duration were valid, then the SNO sensitivity would
also be of order 10 eV.

\begin{figure}[t]
\centerline{\epsfxsize=4in \epsfbox{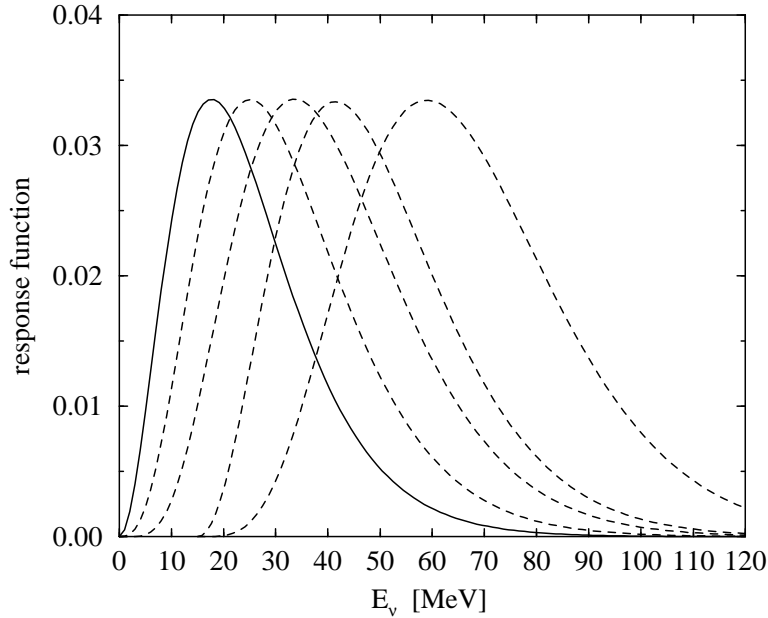}}
\caption{The {\it shapes} of response functions for different
reactions.  The solid line is a Fermi-Dirac distribution $f(E_\nu)$
with temperature 8 MeV.  The dashed lines are products of that with
the various neutral-current cross sections, i.e., $f(E_\nu)
\sigma(E_\nu)$.  The normalizations were chosen so that all curves
would have the same height.  From left to right, the cross sections
are $\nu + e^-$, $\nu + d$, $\nu + ^{12}{\rm C}$, $\nu + ^{16}{\rm
O}$.  For $^{12}$C, only the level at 15.11 MeV was used.  For
$^{16}$O, a fit~\protect\cite{SKpaper} to the full
calculation~\protect\cite{oxy} was used.  In a given experiment, only
the integral of one of these curves is measured, but each samples the
underlying $f(E_\nu)$ differently.  The peak positions are what is
called $E_c$ in the text.}
\end{figure}


\subsection{How Model-Dependent are the Results?}

Above, I discussed the proposed $\langle t \rangle$ technique for
analyzing the supernova neutrino data for a $\nu_\tau$ mass.  This
technique was designed to be as model-independent as possible.  One
underlying theme is to calibrate the assumed models as much as
possible from the data~\cite{SKpaper,SNOpaper}.

While is is probably a good approximation that the luminosities of the
different flavors are equal as a function of time, the normalization
of the luminosity drops out in the definition of $\langle t \rangle$,
and hence the results depend on the weaker assumption that the {\it
shapes} of the luminosities are approximately equal as a function of
time.  The differences among the different flavors at the very early
stages (for example, the higher $\nu_e$ flux) persist over a duration
much less than 10 s, and hence have no effect on the analysis.

Another very important point is that the details of the shape of the
luminosity as a function of time are {\it never used}, because the
delay is measured just from the $\langle t \rangle$ of the
neutral-current events minus the same for the charged-current events.
Why is this enough?  In general, one could describe the scattering
rates of charged-current and neutral-current events by their
normalizations and a series of all moments ($\langle t \rangle$,
$\langle t^2 \rangle$, etc.).  While this would likely require many
terms, it {\it is} efficient to consider the {\it difference} in the
moments between the two.  The mass effects do not change the
normalizations (the numbers of events), and we have shown that the
higher moments (e.g., the width) are not significantly changed by a
mass.  That is, the only statistically significant effects of a mass
on the event rate difference are in the average arrival time $\langle
t \rangle$.  If the details of the supernova model are not known from
theory, then $\langle t \rangle$ is what is called a sufficient
statistic for the mass~\cite{stats}.

The only aspect of the shape of the luminosity $L(t)$ that was used in
the analysis was its duration, and this appears only in the error on
the delay.  {\it Thus the particular assumption of using an
exponential luminosity completely disappears from the problem}.  Any
other reasonable model of similar duration would give a very similar
result.

One also needs to know the $\nu_\tau$ temperature to determine the
mass limit.  Since none of the neutral-current reactions measure the
incoming neutrino energy, there is no direct spectral information.
The $\nu_\tau$ temperature must be determined from the number of
events.  Since different neutral-current reactions have different
energy dependence, one can crudely reconstruct the underlying spectrum
(see Fig.~3).  The moments analysis of the rates reveals that only the
average delay is significant, and thus that just a single energy (the
characteristic energy $E_c$) is relevant, instead of the fine details
of the spectral shape.

The final results for the mass sensitivity are given in Table~III.


\newpage

\section{Supernova Location by Neutrinos}


\subsection{Overview}

There has been great interest recently in the question of whether or
not a supernova can be located by its neutrinos.  If so, this may
offer an opportunity to give an early warning to the astronomical
community, so that the supernova light curves can be observed from the
earliest possible time.  An international supernova early alert
network has been formed for this purpose, and the details of its
implementation~\cite{Habig,nu98} were the subject of a recent
workshop~\cite{SNwksp}.  One of the primary motivations for such a
network is to greatly reduce the false signal rate by demanding a
coincidence between several different detectors.  The second
motivation is to locate the supernova by its neutrino signal.

The interest in the latter is driven by two considerations.  First,
the neutrinos leave the core at the time of collapse, while the
electromagnetic radiation leaves the envelope some hours later,
depending on the stellar mass.  Thus the neutrino observations can
give the astronomers a head start.  Second, the next Galactic
supernova will likely be in the disk, and hence obscured by dust, and
perhaps never visible optically.

There are two classes of techniques to locate a supernova by its
neutrinos.  The first class of techniques is based on angular
distributions of the neutrino reaction products, which can be
correlated with the neutrino direction.  In this case, a single
experiment can independently announce a direction and its error.  The
second method of supernova location, triangulation, is based on the
arrival-time differences of the supernova pulse at two or more
widely-separated detectors.  This technique would require significant
and immediate data sharing among the different experiments.


\subsection{Angular Distributions}

\subsubsection{Neutrino-Electron Scattering: Electron Angular
Distribution}

In neutrino-electron scattering, the scattered electrons have a very
forward angular distribution, due to the small electron mass.  At
supernova neutrino energies, the angle between the incoming neutrino
and the outgoing electron is about $10^\circ$, depending somewhat on
neutrino energy and flavor.  However, multiple scattering of the
struck electron smears out its \v{C}erenkov cone, washing out the
dependence on energy and flavor, and introducing an angular resolution
of about $25^\circ$.

Naively, if the one-sigma width of the electron angular distribution
is $25^\circ$, then the precision with which its center (i.e.,
the average) can be defined given $N_S$ events is
\begin{equation}
\delta \theta \simeq \frac{25^\circ}{\sqrt{N_S}}\,,
\end{equation}
where $N_S$ is the number of events.  For SK~\cite{SKpaper}, $N_S
\simeq 320$, so the cone center could be defined to within about
$1.5^\circ$.  For SNO~\cite{SNOpaper} (using both the light and heavy
water), $N_S \simeq 25$, so the cone center could be defined to about
$5^\circ$.  The equivalent error on the cosine is $\delta(\cos\theta)
\simeq (\delta\theta)^2/2$, i.e., $3 \times 10^{-4}$ and $4 \times
10^{-3}$, respectively.

However, the neutrino-electron scattering events are indistinguishable
on an event-by-event basis from other reactions with only a single
electron or positron detected.  The primary background in SK or the
light water in SNO is thus $\bar{\nu}_e + p \rightarrow e^+ + n$ (in
the heavy water in SNO, there are several background reactions).

This problem is similar to the SK solar neutrino studies, in which the
neutrino-electron scattering events have to be separated from the
intrinsic detector background (due to radioactivity, etc).  The SK
results are presented as an angular distribution in the cosine along
the known direction to the Sun, and clear bump in the solar direction
is seen, with signal/noise $\simeq 1/10$.  Though the concept is the
same, for the supernova events the signal/noise is only $\simeq 1/30$.
Taking this into account~\cite{SNpointing} makes the centroiding error
larger by a correction factor of $\simeq 4$ for both SK and SNO.  With
some cuts on energy, it may be possible to reduce this correction
factor to $\simeq 2 - 3$.

\subsubsection{Neutrino-Nucleus Scattering: Lepton Angular
Distributions}

In the charged-current reactions of $\nu_e$ and $\bar{\nu}_e$ on
nucleons or nuclei, the outgoing electrons or positrons typically
have angular distributions of the form
\begin{equation}
\frac{{\rm d}\sigma}{{\rm d }\cos\theta}
\simeq 1 + v_e a(E_\nu)\cos\theta\,,
\label{eq:angdist}
\end{equation}
where $\theta$ is the angle between the neutrino and electron (or
antineutrino and positron) directions in the laboratory (where the
nuclear target is assumed to be at rest) and $v_e$ is the lepton
velocity in $c = 1$ units.  At energies higher than for supernova
neutrinos, terms proportional to higher powers of $\cos\theta$ appear,
and at the highest energies, all reaction products are strongly
forward simply by kinematics.  Just above threshold, where the lepton
velocity is small, the angular distribution obviously becomes
isotropic.  At supernova neutrino energies, $v_e = 1$.

Typically, the asymmetry coefficient $a(E_\nu)$ is not large, so the
angular distribution is weak, but the number of events may be large
(for $\bar{\nu}_e + p \rightarrow e^+ + n$, there are nearly $10^4$
events expected in SK).  Thus it may be possible to use these events
to locate the supernova.  We first consider how well one could
localize the supernova assuming that $a$ is known and constant.

Given a sample of events, one can attempt to find the axis defined by
the neutrino direction.  Along this axis, the distribution should be
flat in the azimuthal angle $\phi$ and should have the form in
Eq.~(\ref{eq:angdist}) in $\cos\theta$.  Along any other axis, the
distribution will be a complicated function of both the altitude and
azimuthal angles.  We assume that the axis has been found numerically,
and ask how well the statistics allow the axis to be defined.  A
convenient way to assess that is to define the forward-backward
asymmetry as
\begin{equation}
A_{FB} = \frac{N_F - N_B}{N_F + N_B}\,,
\end{equation}
where $N_F$ and $N_B$ are the numbers of events in the forward and
backward hemispheres.  The total number of events is $N = N_F + N_B$.
Note that $A_{FB}$ will assume an extremal value $A_{FB}^{extr}$ along
the correct neutrino direction.  It can be shown~\cite{SNpointing}
that
\begin{equation}
\delta A_{FB}
= \frac{1}{\sqrt{N}} \sqrt{1 - \left(\frac{a}{2}\right)^2}
\simeq \frac{1}{\sqrt{N}}\,,
\end{equation}
where the error is nearly independent of $a$ for small $|a|$, which is
the case for the reactions under consideration.

In the above, the coordinate system axis was considered to be
correctly aligned with the neutrino direction.  Now consider what
would happen if the coordinate system were misaligned.  While in
general, all three Euler angles would be needed to specify an
arbitrary change in the coordinate system, symmetry considerations
dictate that the computed value of $A_{FB}$ depends only upon one --
the angle $\theta$ between the true and the supposed neutrino axis.
Thus $A_{FB}$ is some function of $\theta$ if the axis is misaligned.
Using a Legendre expansion, one can show that
\begin{equation}
A_{FB}(\theta) = \frac{a}{2} \cos\theta\,.
\label{eq:Afb}
\end{equation}
The error on the alignment is then
\begin{equation}
\delta (\cos\theta) =
\frac{2}{|a|} \, \delta A_{FB} \simeq
\frac{2}{|a|} \frac{1}{\sqrt{N}}\,.
\end{equation}

Treating the nucleons as infinitely heavy, the coefficient $a$ in Eq.
(\ref{eq:angdist}) is related to the competition of the Fermi (no spin
flip) and Gamow-Teller (spin flip) parts of the matrix element
squared:
\begin{equation}
a = \frac{ |M_F|^2 - |M_{GT}|^2}{ |M_F|^2 + 3 |M_{GT}|^2} \,.
\end{equation}
Naive use of this formula for $\bar{\nu}_e + p \rightarrow e^+ + n$
gives $a = -0.10$ for ($|M_{GT}/M_F| = 1.26$).  One expects $10^4$
events in SK, so that $\delta (\cos\theta) \simeq 0.2$.  Even though
the asymmetry parameter $a$ is quite small, the number of events is
large enough that this technique in SK could give a reasonable
pointing error (SNO only has 400 of these events, so $\delta
(\cos\theta) \simeq 1.0$, which is too large to be useful).  For the
charged-current reactions on deuterons, $\bar{\nu}_e + d \rightarrow
e^+ + n + n$ and $\nu_e + d \rightarrow e^- + p + p$, one would have
$a = -1/3$ ($M_F/M_{GT} \simeq 0$).  If these channels could be
combined (160 events in total), then $\delta (\cos\theta) \simeq 0.5$,
which is again rather large.

\begin{figure}[t]
\centerline{\epsfxsize=4in \epsfbox{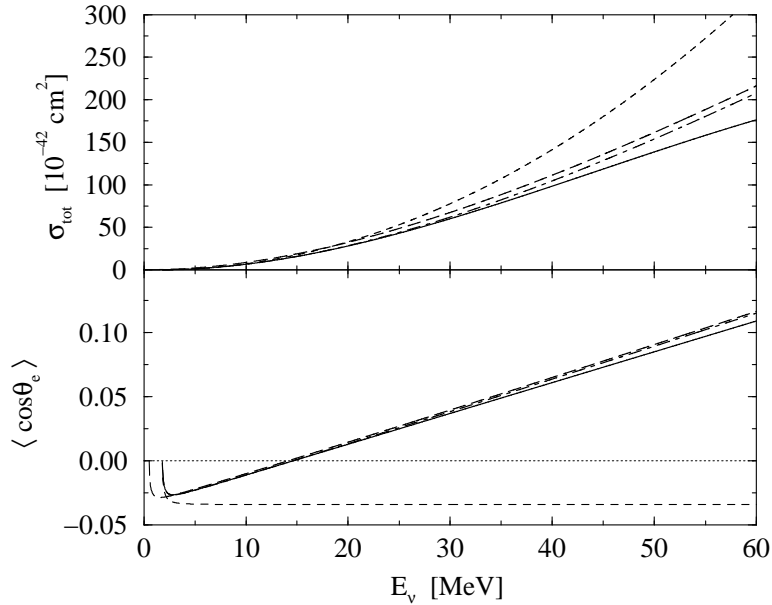}}
\caption{Upper panel: total cross section for $\bar{\nu}_e + p
\rightarrow e^+ + n$; bottom panel: $\langle \cos\theta \rangle$ for
the same reaction; both as a function of the antineutrino energy.  The
solid line is our ${\cal O}(1/M)$ result and the short-dashed line is
the ${\cal O}(1)$ result.  The long-dashed line is the high-energy
formula of Llewellyn-Smith, and the dot-dashed line contains our
threshold modifications to the same.  The inner radiative corrections
are included (see the text), but the outer radiative corrections are
not.  The long-dashed and dot-dashed lines are nearly indistinguishable
in the lower panel.}
\end{figure}

In general, the coefficient $a(E_\nu)$ has energy dependence coming
from recoil and weak magnetism corrections, each of order $1/M$, where
$M$ is the nucleon mass (such terms also change the total cross
section~\cite{Vogel,Fayans,invbeta}, which has some important
applications~\cite{otherWM}).  For the reaction $\bar{\nu}_e + p
\rightarrow e^+ + n$, these corrections have a dramatic effect on the
angular distribution, making it backward at low energies, isotropic at
about 15 MeV, and forward at higher energies.  Averaged over the
expected $\bar{\nu}_e$ spectrum from a supernova, one obtains
$\langle \cos\theta \rangle = +0.08$, i.e., a forward distribution.
Using the common assumption that the angular distribution is backward
(with $a = -0.10$) would lead one to infer {\it exactly the wrong
direction to the supernova}.

It is convenient to describe the angular distribution by the average
cosine, weighted by the differential cross section.  In the limit that
Eq.~(\ref{eq:angdist}) holds,
\begin{equation}
\langle \cos\theta \rangle = \frac{1}{3} v_e a(E_\nu)\,.
\end{equation}
For $\bar{\nu}_e + p \rightarrow e^+ + n$, the variation of
the average cosine can be written as (keeping only the largest
terms of the full expression~\cite{invbeta}):
\begin{equation}
\langle \cos\theta \rangle
\simeq -0.034 v_e + 2.4 \frac{E_\nu}{M}\,.
\label{eq:avgcos1a}
\end{equation}
The factor 2.4 is actually $1.0 + 1.4$, the first term from recoil,
which always makes the angular distribution more forward, and the
second from weak magnetism (which would change sign for the reaction
$\nu_e + n \rightarrow e^- + p$, if there were free neutron targets).
The variation of $\langle \cos\theta \rangle$ with $E_\nu$ is shown
in Fig.~4.

Since the deuteron is weakly bound, the recoil and weak magnetism
corrections to $\langle \cos\theta \rangle$ can be reasonably
estimated~\cite{invbeta}.  The effects of recoil and weak magnetism
add in the $\bar{\nu}_e + d$ channel, and partially cancel in the
$\nu_e + d$ channel, as seen in Fig.~5.

\begin{figure}[t]
\centerline{\epsfxsize=4in \epsfbox{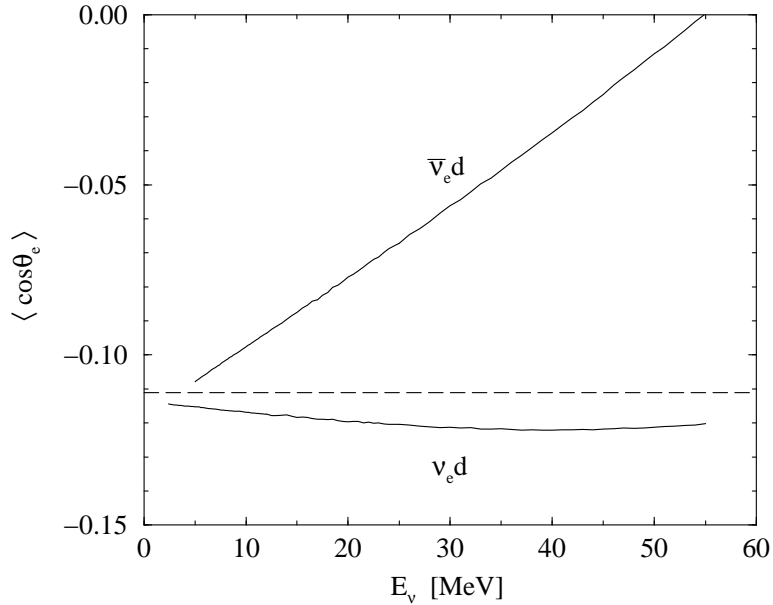}}
\caption{The average lepton cosine for the charged-current deuteron
reactions, versus the neutrino or antineutrino energy, using
Kubodera's calculations.  We plot only from 1 MeV above threshold, so
the $v_e$-dependence at low energies is not shown.  The small jitters
are due to the coarse integration grid.}
\end{figure}

\subsubsection{Inverse Beta Decay: Positron-Neutron Separation Vector}

The results above were based on lepton angular distributions as
observed in water-\v{C}erenkov detectors (light or heavy water).  Some
of the same reactions occur in scintillator detectors, but the lepton
angular distributions are not observable due to the isotropic
character of scintillation light.

Despite that, there is a way to get directional information with a
scintillator detector.  The idea~\cite{Bemporad} makes use of the fact
that scintillator detectors can measure {\it final particle positions}
by the relative timing between different phototubes.  For $\bar{\nu}_e
+ p \rightarrow e^+ + n$, the positron is detected nearly at the point
of creation.  The neutron, however, will be detected (by its capture
gamma rays), on average a few centimeters {\it forward} of the point
of creation.  The initially forward motion of the neutron is just a
consequence of kinematics.  Thus the positron-neutron separation
vector points in the incoming antineutrino direction.

However, before the neutron can be captured, it must be thermalized.
On the first several scatterings (about half of the kinetic energy is
lost on each step), the forward motion tends to be preserved.  In the
remaining scatterings until thermal energy is reached, the scattering
is isotropic.  In the last phase, the neutron wanders randomly,
undergoing further scatterings at thermal energy, until it is
captured.  The neutron wander has the effect of degrading the
significance of the initially forward motion.  The average
displacement and its fluctuations can be calculated by Monte Carlo
simulation~\cite{invbeta,Choozpointing}.

At the energies of reactor antineutrinos, the positron-neutron
separation is about 1.5 cm and the statistical fluctuation several cm,
depending on the neutron capture time in the
scintillator~\cite{invbeta}.  There is an additional uncertainty in
the position due to the gamma-ray localization error.  For reactor
antineutrinos, the source direction is known, so a measurement of the
average positron-neutron separation vector can be used to make a
background measurement, since background events will degrade the
significance of the separation vector from the expectation.  The
precision with which the reactor can be located was evaluated by the
Chooz collaboration and found to be~\cite{Choozpointing} about
$18^\circ$.  Given the small displacement and the large error in
localization, this is an impressive measurement.

At the higher energies of supernova neutrinos, the neutron kinetic
energy is larger and the elastic scattering cross section smaller, so
the displacement is larger.  The Chooz collaboration
estimates~\cite{Choozpointing} that if they were to scale their
detector to the size of SK (32 kton), that they would be able to
locate a supernova to within about $9^\circ$, not so much worse than
the neutrino-electron scattering result of about $5^\circ$ in SK.  The
positron-neutron displacement technique is not possible in SK since
neutrons are not detected, and not possible in SNO because the neutron
wanders over several meters.


\subsection{Triangulation}

Finally, there is the technique of triangulation by arrival-time
differences.  The idea is very simple -- for two detectors, the cosine
of the angle $\theta$ between the axis connecting the two detectors
(with separation $d$) and the supernova direction is determined by the
arrival-time difference $\Delta t$ by
\begin{equation}
\cos\theta = \frac{\Delta t}{d}\,.
\end{equation}
For detectors on opposite sides of the Earth, $d = 40$ ms, and for SK
and SNO, $d = 30$ ms.  Two detectors thus define a cone of allowed
directions to the supernova.  Since the arrival time difference has an
error $\delta(\Delta t)$, the cone will have an angular thickness
\begin{equation}
\delta(\cos\theta) = \frac{\delta(\Delta t)}{d}\,.
\end{equation}
The timing error is obviously the crucial point, and in order to have
$\delta(\cos\theta) = 0.1$ (comparable to the neutrino-electron
scattering result in SNO), one would need $\delta(\Delta t) = 3$ ms
for the SK-SNO difference.  This is much smaller than the duration
of the supernova pulse (10 s) or even the risetime (assumed 100 ms).

In order to assess the best that triangulation could do, consider the
scenario where SK and the light water portion of SNO are each perfect
detectors, identical except for size.  (The shape of the event rate in
the heavy water in SNO will be different, due to the different
reactions, and it is not clear how to correct for that).  Then the
scattering rates observed in SK and the light water in SNO will have
the {\it same underlying distribution}, which depends on unknown
details of the supernova models.  

The observed scattering rates in SK and the light water of SNO can
then differ only in normalization, statistical fluctuations, and a
possible delay.  The normalizations matter only in that they determine
the scale of the statistical fluctuations.  SK will have about $10^4$
events and the light water in SNO about 400 events, so the statistical
error will be dominated by the timing error in SNO.  That is, the
underlying distribution will be measured in SK and this template will
be used in SNO to test for a delay.

In order to make estimates of the timing error, we need the shape of
the event rate.  We assume a short rise (in the current supernova
models, this is somewhere between 0 and 200 ms, so we will very
conservatively use 100 ms), followed by a long decay (10 s, as for
SN1987A).

For the normalized event rate, we take
\begin{eqnarray}
f(t,t_0) & = &
\alpha_1 \times \frac{1}{\tau_1}
\exp\left[+\frac{(t - t_0)}{\tau_1}\right],
\phantom{xxx} t < t_0
\\
f(t,t_0) & = &
\alpha_2 \times \frac{1}{\tau_2}
\exp\left[-\frac{(t - t_0)}{\tau_2}\right],
\phantom{xxx} t > t_0\,.
\end{eqnarray}
where
\begin{equation}
\alpha_1 = \frac{\tau_1}{\tau_1 + \tau_2},
\phantom{xxxxxx}
\alpha_2 = \frac{\tau_2}{\tau_1 + \tau_2}\,.
\end{equation}
Then $f(t,t_0)$ is a normalized probability density function built out
of two exponentials, and joined continuously at $t = t_0$.  In what
follows, we assume that this form of $f(t,t_0)$ is known to be correct
and that $\tau_1$ and $\tau_2$ are {\it known}.  The shape of the
event rate is shown in Fig.~6.

The event rate in SNO will consist of $N$ events sampled from
$f(t,t_0)$, and the event rate in SK will consist of $N'$ events
sampled from $f(t,t_0')$.  Then $\Delta t = t_0 - t_0'$, and $\delta
(\Delta t) \simeq \delta t_0$, since the SNO error dominates.  We
consider only the statistical error determined by the number of
counts.  As noted, we want to determine the {\it minimal} error on the
triangulation.  This model, while simple, contains the essential
timescales and an adjustable offset.

These considerations lead to a well-posed statistical problem: If $N$
events are sampled from a {\it known} distribution $f(t,t_0)$, how
well can $t_0$ be determined?  The Rao-Cramer
theorem~\cite{stats} provides an answer to this question
(with a possible subtlety~\cite{Cousins}).  This
theorem allows one to calculate the minimum possible variance on the
determination of a parameter (here $t_0$), by any technique
whatsoever.  This minimum variance can be achieved when all of the
data are used as ``efficiently'' as possible, which is frequently
possible in practice.  

\begin{figure}[t]
\centerline{\epsfxsize=4in \epsfbox{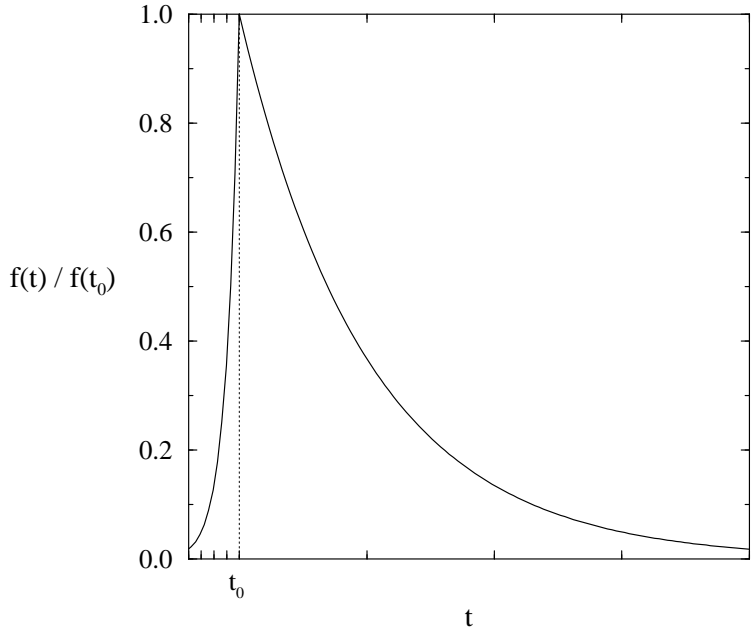}}
\caption{The schematic form of the normalized event rate $f(t,t_0)$ is
shown.  To the left of $t_0$ there is an exponential rise with time
constant $\tau_1$.  To the right of $t_0$ there is an exponential
decay with time constant $\tau_2$.  The tick marks on the $t$-axis are
in units of the respective time constants $\tau_1$ and $\tau_2$.  For
clarity of display, we used $\tau_1/\tau_2 \simeq 10^{-1}$ in the
figure, instead of the $\tau_1/\tau_2 \simeq 10^{-2}$ assumed in the
analysis.}
\end{figure}

One requirement of the theorem is that the domain of positive
probability must be independent of the parameter to be determined.
This condition is obviously not met for a zero risetime, since then
the domain is $(t_0,\infty)$.  For a nonzero risetime, the domain is
technically $(-\infty,\infty)$, independent of $t_0$, and so the
theorem applies.  The minimum possible variance on the determination
of $t_0$ is:
\begin{equation}
\frac{1}{\left( \delta t_0 \right)^2_{\rm min}} = 
N \times
\int dt \, f(t,t_0)
\left[\frac{\partial \ln f(t,t_0)}{\partial t_0}\right]^2\,.
\end{equation}
This is the general form for an arbitrary parameter $t_0$.  When $t_0$
is a translation parameter, i.e., $f(t,t_0)$ depends only on $t -
t_0$, this reduces to
\begin{eqnarray}
\frac{1}{\left( \delta t_0 \right)^2_{\rm min}} & = & 
N \times
\int dt \, f(t,t_0)
\left[\frac{\partial \ln f(t,t_0)}{\partial t}\right]^2 \\
& = & 
N \times
\int dt \, 
\frac{\left[\partial f(t,t_0)/\partial t\right]^2}{f(t,t_0)}\,.
\end{eqnarray}
For the particular choice of $f(t,t_0)$ above, this reduces to
\begin{equation}
\frac{1}{\left( \delta t_0 \right)^2_{\rm min}} = 
N \times
\left(\alpha_1/\tau_1^2 + \alpha_2/\tau_2^2\right)\,,
\end{equation}
and so the minimum error is
\begin{equation}
\left(\delta t_0\right)_{\rm min} =
\frac{\sqrt{\tau_1 \tau_2}}{\sqrt{N}} =
\frac{\tau_1}{\sqrt{N_1}}\,.
\end{equation}
Note that $N_1 = N (\tau_1/\tau_2)$ is approximately the number of
events in the rising part of the pulse.  Since the rise is the
sharpest feature in $f(t,t_0)$, it is unsurprising that it contains
almost all of the information about $t_0$.  The total number of events
$N$ is fixed by the supernova binding energy release, so a change in
the assumed total duration of the pulse, i.e., $\tau_2$, would affect
the peak event rate and hence the fraction of events in the leading
edge.  For a more general $f(t,t_0)$, one would replace
$\tau_1/\tau_2$ by this fraction computed directly.

For SNO, $N_1 \simeq 10^{-2} \times 400 \simeq 4$, so $\delta(t_0)
\simeq 30 {\rm\ ms}/\sqrt{4} \simeq 15$ ms.  Since SK has about 25
times more events, the corresponding error would be about 3 ms.
Therefore, the error on the delay is $\delta(\Delta t) \simeq 15$ ms
and $\delta(\cos\theta) \simeq 0.50$ at one sigma.  We have not
specified the method for extracting $t_0$ and hence $\Delta t$ from
the data~\cite{SNpointing}.  That is exactly the point of the
Rao-Cramer theorem -- that one can determine the minimum possible
error without having to try all possible methods.

If the risetime were zero, which seems to be unrealistic, then one can
show~\cite{SNpointing} by application of order statistics that the
error becomes
\begin{equation}
\delta(t_0) = \frac{\tau_2}{N}\,.
\end{equation}
Here $\tau_2/N$ is simply the spacing between events near the peak.
For a more general $f(t,t_0)$, but still with a sharp edge at $t_0$, one
would simply replace $\tau_2$ by $1/f(t_0)$.  In fact, the shape of
$f(t,t_0)$ is irrelevant except for its effect on the peak rate, i.e.,
$f(t_0)$.  So long as $f(t,t_0)$ has a sharp edge and the right total
duration, allowing a more general time dependence would therefore not
change the results significantly.

For the two cases, zero and nonzero risetime, we used different
mathematical techniques.  This may seem like an artificial
distinction, and that these two cases do not naturally limit to each
other.  In particular, it may seem incompatible that in the first case
the error $\sim 1/\sqrt{N}$, while in the second the error $\sim 1/N$.
Further, one obviously cannot take $\tau_1 \rightarrow 0$ in the first
result to obtain the second.  However, it can be
shown~\cite{SNpointing} that the two techniques have disjoint regions
of applicability (as a function of $\tau_1$) and that the formulas
match numerically at the boundary between the two.  The boundary is
the value of $\tau_1$ such that the number of events in the rise is of
order 1, i.e., the edge appears sharp for this or smaller $\tau_1$.

The final results for the pointing errors are given in Table~IV.


\section{Concluding Remarks}

The next Galactic supernova should be a bonanza for neutrino physics
and astrophysics.  Should, that is, provided that we can make sense of
the data.  The difficulty is that there are many unknown aspects of
both the particle physics properties of neutrinos and the astrophysics
of the expected signal.  Since the expected core-collapse supernova
rate in the Galaxy is about 3 per century~\cite{SNrate}, and since
there are no neutrino detectors sensitive to supernovae in distant
galaxies, we will not have the luxury of many observations.  Of
course, there is a worldwide effort to improve the numerical models of
core-collapse supernovae~\cite{SNmodels}.  And besides neutrinos,
there is one other direct probe of the supernova core, and that is
gravitational radiation.  LIGO may be sensitive to supernovae out to
the Virgo Cluster (at rates of order 1 per year~\cite{LIGO}), and its
observations may improve the understanding of stellar collapse.

Since there are uncertainties in the supernova models, the $\langle t
\rangle$ test for the $\nu_\mu$ and $\nu_\tau$ masses was designed to
be as model-independent as possible.  The resulting sensitivities for
either $\nu_\mu$ or $\nu_\tau$ are in Table~III.  If the $\nu_\mu$ and
$\nu_\tau$ are maximally mixed with a small mass difference, then the
mass test is really on $m_2 = m_3$, and the limits in Table~III
improve by a factor of about $\sqrt{2}$.

\begin{table}[b]
\caption{Mass sensitivity for various detectors and neutral-current
reactions.  The mass quoted is the limit which would be placed on
either the $\nu_\mu$ or $\nu_\tau$ mass at 90\% CL if no delay were
seen (equivalently, it is the smallest mass which would cause an
unambiguous delay).  If the $\nu_\mu$ and $\nu_\tau$ masses are
nearly equal, then the sensitivity is improved by $\protect\sqrt{2}$.
See also~\protect\cite{Cadonati} for the $^{12}$C results.}
\vspace{0.2cm}
\begin{center}
\begin{tabular}{|l|l|l|}
\hline
detector & neutral-current signal & mass limit \\
\hline\hline
SNO & $\nu + d$ & 30 eV \\
SK  & $\nu + ^{16}{\rm O}$ & 45 eV \\
SK  & $\nu + e^-$ & 50 eV \\
Kamland & $\nu + ^{12}{\rm C}$ & 55 eV \\
Borexino & $\nu + ^{12}{\rm C}$ & 75 eV \\
\hline
\end{tabular}
\end{center}
\end{table}

There are also many possible signals of supernova neutrino
oscillations~\cite{SNosc}.  Ideally, the current and near-term
terrestrial neutrino oscillation experiments can answer some of the
key questions before the next supernova.  Particularly crucial is the
question of whether there are sterile neutrinos (or several flavors
thereof).  SNO~\cite{SNO} may tell us whether the solar $\nu_e$
neutrinos are oscillating into sterile or active flavors (or at all).
Similarly for SK~\cite{SK}, K2K~\cite{K2K}, and MINOS~\cite{MINOS} and
the atmospheric $\nu_\mu$ neutrinos.  And MiniBoone~\cite{MiniBoone}
may decide whether the LSND~\cite{LSND} anomaly is correct (and hence
whether sterile neutrinos are necessary).  These questions and more
are discussed thoroughly in several recent reviews~\cite{reviews}.

Since the neutrinos leave the supernova core a few hours before the
light leaves the envelope, it should be possible for the neutrino
experiments to give astronomers an advance warning of a Galactic
core-collapse supernova, and also to provide some guidance as to where
in the sky to look.  Neutrino-electron scattering in SK seems to be
the best technique, with an error of order $5^\circ$, depending on
distance and the suppression of backgrounds from other reactions.
While $5^\circ$ may sound large to an astronomer, it is about 0.1\% of
the sky, which will make subsequent searches with small telescopes
much easier (the role of amateur astronomers is discussed in a recent
article~\cite{Leif}).  Supernova location by triangulation seems to be
rather difficult at present (see Table~IV).

\begin{table}[h]
\caption{One-sigma errors on how well the direction to the supernova
is defined by various techniques, at $D = 10$ kpc.  The other parameters
used are noted in the text.  For neutrino-electron scattering, the most
pessimistic background assumptions were used.}
\vspace{0.2cm}
\begin{tabular}{|l|r|}
\hline
technique & error \\
\hline\hline
$\nu + e^-$ forward scattering (SK) &
 $\delta \theta \simeq 5^\circ$,
 $\delta(\cos\theta) \simeq 4 \times 10^{-3}$ \\
$\nu + e^-$ forward scattering (SNO) & 
 $\delta \theta \simeq 20^\circ$,
 $\delta(\cos\theta) \simeq 6 \times 10^{-2}$ \\
\hline
$\bar{\nu}_e + p$ angular distribution (SK) &
 $\delta(\cos\theta) \simeq 0.2$ \\
$\bar{\nu}_e + p$ angular distribution (SNO) &
 $\delta(\cos\theta) \simeq 1.0$ \\
$\nu_e + d, \bar{\nu}_e + d$ angular distributions (SNO) &
 $\delta(\cos\theta) \simeq 0.5$ \\
\hline
triangulation (SK and SNO) &
 $\delta(\cos\theta) \simeq 0.5$ \\
\hline
\end{tabular}
\end{table}


\newpage

\section*{ACKNOWLEDGMENTS}

I am grateful to Gabor Domokos and Susan Kovesi-Domokos for the
invitation to a very interesting workshop, and to Petr Vogel for his
collaboration~\cite{SKpaper,SNOpaper,SNpointing,invbeta}.  This work
was supported by a Sherman Fairchild fellowship at Caltech.




\begin{thebibliography}{99}
\bibitem{Cousins}
Bob Cousins (UCLA) reminds me that what we really want is the
minimum-sized confidence interval, which does not always correspond to
a minimum variance (e.g., the Lorentzian distribution).  However,
there is apparently no such pathology here.
\end{thebibliography}
\end{document}